# Field-tunable toroidal moment and anomalous Hall effect in noncollinear antiferromagnetic Weyl semimetal Co$_{1/3}$TaS$_2$

Pyeongjae Park[1,2], Yoon-Gu Kang[3], Junghyun Kim[1,2], Ki Hoon Lee[4], Han-Jin Noh[5], Myung Joon Han[3*], and Je-Geun Park[1,2,6*]

[1]*Center for Quantum Materials, Seoul National University, Seoul 08826, Republic of Korea*

[2]*Department of Physics & Astronomy, Seoul National University, Seoul 08826, Republic of Korea*

[3]*Department of Physics, KAIST, Daejeon 34141, Republic of Korea*

[4]*Department of Physics, Incheon National University, Incheon, 22012, Republic of Korea*

[5]*Department of Physics, Chonnam National University, Gwangju 61186, Republic of Korea*

[6]*Institute of Applied Physics, Seoul National University, Seoul 08826, Republic of Korea*

\* Corresponding author: jgpark10@snu.ac.kr & mj.han@kaist.ac.kr

## Abstract

Combining magnetism with band topology provides various novel phases that are otherwise impossible. Among several cases, noncollinear metallic antiferromagnets can reveal particularly rich topological physics due to their diverse magnetic ground states. However, there are only a few experimental studies due to the lack of suitable materials, especially with triangular lattice antiferromagnets. Here, we report that metallic triangular antiferromagnet Co$_{1/3}$TaS$_2$ exhibits a substantial anomalous Hall effect (AHE) related to its noncollinear magnetic order. Our first-principles calculations found that hourglass Weyl fermions from the non-symmorphic symmetry trigger AHE. We further show that AHE in Co$_{1/3}$TaS$_2$ can be characterized by the *toroidal moment*, a vortex-like multipole component that arises from a combination of chiral lattice and geometrical frustration. Finally, the unusual field-tunability of the toroidal moment puts Co$_{1/3}$TaS$_2$ as a rare example of a noncollinear metallic antiferromagnet filled with interesting magnetic and topological properties.

**Introduction**

Introducing magnetic order into electron band structure presents anexciting opportunity to realize topological phases otherwise unavailable in non-magnetic systems. Such phases can emerge from perturbations due to magnetism on the electronic band structure's geometry and topology. Anomalous Hall effect (AHE) in metallic ferromagnets is the most direct consequence of such perturbations. However, the more interesting case of magnetic topological band structure lies with antiferromagnets, which provide several interesting situations and could, in principle, offer unseen properties arising simply from the enormous number of possible spin configurations[1,2]. One of the most remarkable properties is the sizable AHE in a metallic antiferromagnet despite its zero or vanishingly small net magnetization, which is roughly proportional to AHE in the case of ferromagnets. Thus, a different organizing principle than magnetization on its own should govern the AHE in antiferromagnets.

To understand AHE in metallic antiferromagnets, one should identify an effective order parameter that characterizes the symmetry-breaking field that leads to finite AHE. One example is AHE from non-coplanar magnetic order (also known as topological Hall effect), which can be well characterized by scalar spin chirality[3,4]. More interestingly, it has been recently suggested that noncollinear antiferromagnets with coplanar magnetic order can also manifest a large AHE with the aid of spin-orbit coupling[5-9]. In this case, AHE can be characterized by the cluster multipole moment[10,11]. Despite such an exciting organizing principle, however, there are still no relevant experimental results except for the only one known example of such: that is, the cluster octupole moments in $Mn_3X$ (X=Ge, Sn, Ir)[1,2,10-12]. In other words, the richness of noncollinear antiferromagnetic configuration space with non-trivial band structure is just a hypothetical proposition at the moment. Hence, finding another noncollinear antiferromagnet is the crucial way forward to materialize the untapped potential of antiferromagnetic metallic systems, which would give access to AHE originating from a much more diverse multipole magnetism, where unique physics may be found in each case. This may, in particular, hold for a multipole called toroidal moment. Because it requires breaking both space inversion and time-reversal symmetry (TRS), the ordering of toroidal moments (i.e., ferro-toroidal order) is distinct from that of magnetic multipoles[13] and therefore can manifest some non-trivial phenomena. However, due to the scarcity of suitable materials, relatively less attention has been paid to this multipole class so far, which prevents a complete understanding of its connection to macroscopic quantities measured by experiments.

$TM_{1/3}MS_2$ (TM=3d transition metals and M=Ta, Nb) are promising materials for studying metallic antiferromagnetism in a triangular lattice[14,15]. It consists of well-known metallic $2H$-$MS_2$ with triangular layers of TMs intercalating into a vdW gap of $2H$-$MS_2$ (Fig. 1a). Such intercalation slightly alters the atomic positions of sulfur atoms, leading to the chiral structure with a non-centrosymmetric $P6_322$ space group (No. 182)[16]. The charge transfer from TM to $2H$-$MS_2$ forms divalent ($TM^{2+}$) or trivalent ($TM^{3+}$) TM ions, and subsequently localized magnetic moments settle down in TM sites (Fig. 1b). Interestingly, this material class exhibits various magnetic ground states depending on the choice of TM and M[14]. Among the combinations, $Co_{1/3}TaS_2$ is the only material with the noncollinear 120° magnetic order[17], a natural outcome due to geometrical frustration (Fig. 1b). Therefore, $Co_{1/3}TaS_2$ offers a combination of chirality, triangular lattice, and 120° magnetic structure, unique among the current set of magnetic topological materials. Moreover, $TM_{1/3}MS_2$ can be thinned down to a few atomic layers through mechanical exfoliation[18]. Hence, it could also open up an entirely new route to study noncollinear antiferromagnetism and corresponding topological physics in genuine two-dimension.

This work reports that single-crystal $Co_{1/3}TaS_2$ is a metallic triangular antiferromagnet where coplanar 120° magnetic order plays a crucial role in forming electron band topology and concomitant AHE. Our density functional theory (DFT) calculations show that AHE is triggered by hourglass Weyl fermions enforced by non-symmorphic symmetry under broken TRS. Moreover, we explain how the toroidal moment, a multipole component in non-centrosymmetric chiral antiferromagnets[13], can characterize AHE in $Co_{1/3}TaS_2$. Finally, we discuss the remarkable tunability of the toroidal moment under an out-of-plane magnetic field, which implies a profound correlation between AHE and the magnetism of $Co_{1/3}TaS_2$.

**Results and Discussion**

**Bulk properties** Temperature-dependent magnetization (**M**), resistivity ($\rho_{xx}$), and specific heat ($C_p$) of $Co_{1/3}TaS_2$ indicate two magnetic phase transitions at $T_{N1}$=38 K and $T_{N2}$=26.5 K (Fig. 1c-f). The major transition occurs at $T_{N2}$, where all the properties change significantly. The negative Curie-Weiss temperature ($\theta_{CW}$) with a significant frustration factor $f \equiv \frac{|\theta_{CW}|}{T_{N2}}$ ~4 indicates dominant antiferromagnetic interactions with frustration, consistent with the formation of the 120° order. Interestingly, the phase below $T_{N2}$ involves a weak ferromagnetic moment of 0.01 $\mu_B$/$Co^{2+}$ along the c-axis ($M_z(\mathbf{H} = 0)$), as shown in the field dependence of $M_z$ with hysteresis (Fig. 1d). In addition to the domain switch of $M_z(\mathbf{H} = 0)$ at $\pm H_{c1}$, there is a jump at $\pm H_{c2}$ due to a meta-magnetic transition. The mechanism of this meta-magnetic transition remains unclear due to limited information, which requires further studies.

Most interestingly, the AHE ($\rho_{xy}(\mathbf{H} = 0)$) appears below $T_{N2}$, despite the antiferromagnetic nature of $Co_{1/3}TaS_2$ (Fig. 2a). The anomalous Hall conductivity (AHC, $\sigma_{xy}(\mathbf{H} = 0)$) reaches a value of 70 $\Omega^{-1}cm^{-1}$ at 2 K, comparable to those measured in ferromagnetic materials[3]. After excluding the linear Hall effect, the magnitude of the $\rho_{xy}(\mathbf{H})$ remains nearly unchanged with varying a magnetic field, except for a jump at $H_{c2}$. Also, the observed hysteresis of AHE with a significant coercive field $H_{c1}$ is quite similar to that of the weak ferromagnetic moment (Fig. 2b), particularly in that they simultaneously flip to the opposite domain. This connection between the two quantities shall be discussed later in this paper.

The shape of the hysteresis itself is interesting and merits separate comments. First of all, the meta-magnetic transition at $\pm H_{c2}$ already starts to appear when $T_{N2}<T<T_{N1}$ (Supplementary Figure 5), indicating that the meta-magnetic transition is more related to the magnetic order corresponding to $T_{N1}$. Second, the domain switch at $H_{c1}$ and the meta-magnetic transition at $H_{c2}$ are not independent. While $H_{c2}$ remains nearly unchanged, $H_{c1}$ increases rapidly with decreasing temperature from $T_{N2}$ and reaches almost 9 T at 2 K (Fig. 2c). Indeed, the crossover between $H_{c1}$ and $H_{c2}$ should occur somewhere below $T_{N2}$, around 14 K (Fig. 2c). Surprisingly, the domain switch and the meta-magnetic transition happen simultaneously at $H_{c1}$ after the crossover ($|H_{c1}|>|H_{c2}|$), resulting in a peculiar shape of hysteresis below 15 K (Fig. 2a-b). Thus, the two-fold magnetic domain with respect to the c-axis, the metamagnetic transition at $\pm H_{c2}$, and the magnetic order that forms at $T_{N1}$ and $T_{N2}$ are deeply related to each other, which requires further studies for a complete understanding.

Moreover, the temperature evolution of $\sigma_{xy}(\mathbf{H}=0)$ in $Co_{1/3}TaS_2$ behaves just like an order parameter of the phase transition at $T_{N2}$ (Fig. 2d), implying that the observed AHE is closely related to the magnetic order of $Co_{1/3}TaS_2$. This hypothesis is further supported by the fact that the critical exponent $\beta$ of $\sigma_{xy}(T,\mathbf{H}=0)$ is nearly identical to that of $M_z(T,\mathbf{H}=0)$, as shown in Fig. 2e. Note that the fitted $\beta$ is close to 0.5, consistent with the fact that the magnetic order mediated by long-ranged RKKY interaction follows the mean-field critical behavior[19]. Finally, we compared the magnitude of $\sigma_{xy}(\mathbf{H}=0)$ measured below 15 K for several samples with different Co compositions (0.310<x<0.325). While $\rho_{xx}$ and $\rho_{xy}$ vary depending on the samples, $\sigma_{xy}(\mathbf{H}=0)$ and the shape of the hysteresis remain nearly the same for all the samples (Fig. 2f and Supplementary Figure 6).

**Theoretical understanding of AHE** The observed AHC's robustness and its close connection to the magnetic order imply the intrinsic nature of the AHE in $Co_{1/3}TaS_2$. The following two possibilities are not likely the situation of $Co_{1/3}TaS_2$, based on our findings regarding its microscopic relation to the magnetic order. First, although $Co_{1/3}TaS_2$ has a small weak ferromagnetic moment, the observed AHE in $Co_{1/3}TaS_2$ cannot be totally attributed to the weak ferromagnetic moment itself. This is because the field evolution of AHE does not match with that of $M_z$; see $dM_z/dH$ and $d\rho_{xy}/dH$ at $H_{c1}$ and $H_{c2}$. Also, $M_z(\mathbf{H}=0) \sim 0.01\ \mu_B/Co^{2+}$ itself is too small to induce $\sigma_{xy}(\mathbf{H}=0)$ as large as $\sim 70\ \Omega^{-1}cm^{-1}$. Nor could it originate from the scalar spin chirality due to the slight out-of-plane canting, as the spin chirality is usually canceled out in a triangular lattice antiferromagnet with 120° order[20]. Therefore, understanding AHE in $Co_{1/3}TaS_2$ requires a new idea related to the coplanar 120° spin order itself, analogous to the case in noncollinear kagome antiferromagnet $Mn_3X$ (X=Ge, Sn)[5,7].

We first begin with symmetry arguments for the possible 120° magnetic order in $Co_{1/3}TaS_2$, which can formulate the existence of AHE. According to the group-theoretical analysis[21], the possible magnetic ground states can be described by six irreducible representations referred to as $\Gamma_1 \sim \Gamma_6$ (Supplementary Figure 8). Note that $\Gamma_1$, $\Gamma_2$, and $\Gamma_5$ spin configurations can be transformed into each other through uniform spin rotation around the c-axis ($\theta$ in Fig. 3a). Notably, the two-fold rotation ($C_{2a}$ or $C_{2a*}$) and the $6_3$ screw symmetry combined with time-reversal ($\equiv \underline{6_3}$) forbid finite $\sigma_{xy}(\mathbf{H}=0)$ and $M_z(\mathbf{H}=0)$ in $Co_{1/3}TaS_2$: e.g., $\Gamma_1$ ($P6_322$) in Fig. 3a (for more explanation, see Supplementary Text). Consequently, only $\Gamma_2$ ($P6_3\underline{22}$) and the $\Gamma_5$ ($P6_3$) can possess finite $\sigma_{xy}(\mathbf{H}=0)$ and $M_z(\mathbf{H}=0)$ and explain our experimental results. Yet $\Gamma_5$ is a two-dimensional irreducible representation with much lower symmetry, and therefore, is less likely to occur in natural materials, as pointed out in previous studies of two-dimensional triangular antiferromagnets[22]. Thus, we focus on the $\Gamma_2$ spin configuration in our analysis (Fig. 3b and 3c). Our data indeed support such a choice as magnetic susceptibility measured along the a-axis is slightly larger than that along the $a^*$-axis (Fig. 1c); aligning the spins along the a-axis or its symmetrically equivalent directions is energetically more favorable (i.e., local easy-axes), which is the case of $\Gamma_2$ as shown in Fig. 3b. Also, as shown in Fig. 3d, the total energy obtained from our DFT calculation reaches its minimum when $\theta=0°$ ($=\Gamma_2^+$) or 180° ($=\Gamma_2^-$), which is another evidence supporting the $\Gamma_2$ ground state.

The observed intrinsic AHE can be understood as a manifestation of novel Weyl fermions in the electron band structure of $Co_{1/3}TaS_2$, where non-symmorphic symmetry enforces their existence. Notably, the crystal structure of $Co_{1/3}TaS_2$ has the $6_3$ screw symmetry that remains unbroken in the $\Gamma_2$ representation. Such non-symmorphic symmetry could result in specific Weyl crossings in the **k**-space called hourglass Weyl fermion[23-25]. Fig. 4a shows the schematic band dispersion representing a typical situation of hourglass Weyl fermions. Along the screw-invariant line connecting time-reversal invariant momenta (TRIM) points (say, $\Gamma-A$ and $M-L$), there must be band crossings as guaranteed by screw symmetry. By half-translation, the eigenstates of the $2_1$ screw symmetry operation acquire a momentum-dependent phase factor of $e^{-ik_z c/2}$ (c is the lattice constant along the c-axis), and it then causes Kramers' pair exchange (indicated by using color code) as $k_z$ moves from one TRIM point to another. Consequently, the pairs of bands must cross at least once within the invariant space, which naturally makes the band dispersion have the 'hourglass' shape and creates inevitable Weyl crossing (olive circles in Fig. 4a).

To confirm this scenario for $Co_{1/3}TaS_2$, we carried out DFT calculations with the $\Gamma_2$ spin configuration (Fig. 4b): we also tested other possible spin configurations with less promising results. As marked in color, certain pockets of the **k**-space contain the anticipated novel hourglass type dispersion enforced by the $6_3$ screw symmetry. It is especially notable that, protected by the screw symmetry, the Weyl 'crossing' point in the hourglass dispersion can be preserved even under TRS breaking[26]. Fig. 4c and 4d show the enlarged band dispersion along the $\overline{\Gamma A}$ and $\overline{ML}$ lines using the $\Gamma_2$ spin configuration (Fig. 4b), where the color scale indicates the $2_1$ screw rotation eigenvalues. Due to the absence of TRS, $\Gamma$, A, M, and L are no longer TRIM points (Fig. 4c and 4d). Thus, the Kramers' degeneracies are lifted at these points (indicated by gray arrows in Fig. 4d), although the degeneracy lifting is not visible at the $\Gamma$ and A points in this energy scale. However, the hourglass dispersion and the Weyl crossings (the olive circles in Fig. 4c and 4d) are still well maintained. This is clearly shown when comparing Fig. 4c and 4d with Fig. 4a; the evolution of the screw operation's eigenvalue is essentially the same as each other. We note that few theoretical studies have shown the hourglass Weyl fermions with broken TRS, which adopted simple ferromagnetic or collinear antiferromagnetic spin order[27-29].

From the AHC's point of view, however, the Weyl crossings along the $\overline{\Gamma A}$ and $\overline{ML}$ lines, albeit interesting, cannot be responsible for the observed AHE. It is simply because their locations are too far off from the Fermi level (see Fig. 4b). Instead, we found that the Weyl points on the $\overline{KH}$ line play a more critical role. Although both K and H points are not the TRIM points even with TRS, the existence of Weyl points on the $\overline{KH}$ line is guaranteed by the hourglass Weyl points on the $\overline{\Gamma A}$ and $\overline{ML}$ lines due to the fermion doubling theorem[30,31]. From the fact that Weyl points on the $\overline{\Gamma A}$ and $\overline{ML}$ line respectively have a multiplicity of 1 and 3 in half Brillouin zone (BZ), Weyl points are required to lie also on the $\overline{KH}$ line with a multiplicity of 2 to neutralize the chiral charge: $\nu_{\Gamma A} + 3\nu_{ML} + 2\nu_{KH} = 0$. As displayed in Fig. 4e, there are indeed multiple Weyl crossings along the $\overline{KH}$ line, generating Berry curvature like a magnetic monopole. Moreover, since they are located close enough to the Fermi level, they are likely to make a dominant contribution to the anomalous charge transport. The results presented in Fig. 4f constitute more convincing evidence supporting our argument. The black and orange line represents the calculated AHC obtained by integrating the whole BZ and only along the $\overline{KH}$ line, using several 120° spin configurations specified by $\theta$ in Fig. 3a, i.e., using $\Gamma_1$ ($\theta = 90°$ or 270°), $\Gamma_2$ ($\theta = 0°$ or 180°), and $\Gamma_5$ (other angles) spin configurations. First, the calculated $|\sigma_{xy}|$ using the $\varGamma_2$ configuration is about 90 $\Omega^{-1}\text{cm}^{-1}$, which is indeed in excellent agreement with our experimental results (Fig. 2d and 2f). Also, only a small difference is found between the $\sigma_{xy}$ calculated over the entire BZ and that obtained only along the $\overline{KH}$ line, implying that the most dominant contribution of $\sigma_{xy}$ comes from the Weyl points in this particular $\overline{KH}$ symmetry line. Taken together, the 'vestige' of hourglass Weyl fermions on the $\overline{\varGamma A}$ and $\overline{ML}$ lines, which remains intact under the 120° magnetic order, triggers the AHE in this material despite the broken TRS and the lifted degeneracy otherwise residing at Γ, A, M, and L.

**AHE and toroidal moment** Fig. 4f further demonstrates that the AHC strongly depends on the detailed in-plane spin configuration of $Co_{1/3}TaS_2$. To put it intuitively, a vortex-like arrangement inside the 120° spin configuration breaks the $C_{2a}$ or $C_{2a^*}$ symmetry and allows finite $\sigma_{xy}(\mathbf{H}=0)$ in $Co_{1/3}TaS_2$. Fig. 3a and 3b show two clear cases: one is without the vortex-like arrangement of $\varGamma_1$ and another is with the vortex-like arrangement of $\varGamma_2$. Such an arrangement can be precisely quantified by the toroidal dipole moment (**t**)[11,13,32]:

$$\mathbf{t} \equiv \sum_i \mathbf{r}_i \times \mathbf{S}_i \qquad (1)$$

where $\mathbf{r}_i$ is the position vector of the $i^{\text{th}}$ site from the center of the six Co atoms with hexagonal configuration (Fig. 3b). Unlike ordinary magnetic or electric multipoles, **t** breaks both inversion and time-reversal symmetry and consequently appears in a non-centrosymmetric chiral magnet with geometrical frustration[13,32]. Therefore, $Co_{1/3}TaS_2$ is a rare system that can have non-zero **t**, and so it is helpful for a proper understanding of its role in characterizing AHE, which has been suggested only in theory so far[11]. Fig. 3e shows a complete comparison between $t_z$ and calculated $\sigma_{xy}$ for the $\varGamma_1$, $\varGamma_2$, and $\varGamma_5$ spin configurations. As expected, it clearly reveals the aspect of $t_z$ as the order parameter characterizing the AHE.

From this point of view, we can interpret the $\Gamma_2^+$ ($\theta = 0°$) and $\Gamma_2^-$ ($\theta = 180°$) spin configurations as a pair of ferro-toroidal states with an opposite sign of $t_z$ (Fig. 3b and 3c). Interestingly enough, the sign change of $\sigma_{xy}$ at $\pm H_{c1}$ then corresponds to the transition between the $\Gamma_2^+$ and the $\Gamma_2^-$ configurations (Fig. 3f). In other words, an out-of-plane magnetic field switches the AHE of $Co_{1/3}TaS_2$ by manipulating the in-plane spin configuration. While such a consequence may appear at first sight somewhat surprising, it is the weak ferromagnetic moment that connects the out-of-plane field and the in-plane spin configuration in $Co_{1/3}TaS_2$. The simultaneous jump of $\sigma_{xy}$ and $M_z$ seen in our experiments (Fig. 2a and 2b) implies that the sign of $t_z$ is coupled to the sign of $M_z(\mathbf{H} = 0)$; i.e., the $\Gamma_2^+$ ($\Gamma_2^-$) configuration always tends to have positive (negative) $M_z(\mathbf{H} = 0)$. Consequently, the out-of-plane magnetic field can effectively control the sign of $t_z$ by flipping $M_z(\mathbf{H} = 0)$, consistent with the similar hysteretic behavior of $\sigma_{xy}$ to that of $M_z$ (see Fig. 2a and 2b). The coexistence of the toroidal moment and the weak ferromagnetic moment in $Co_{1/3}TaS_2$ is not accidental. A combination of lattice chirality and toroidal spin configuration results in the broken two-fold rotation ($C_{2a}$, $C_{2a^*}$), mirror, and TRS. Such a situation leads to the emergence of a finite vector quantity along the chiral axis (the c-axis in $Co_{1/3}TaS_2$), whose transformation property is equivalent to that of magnetization[33,34]. Of further interest, $t_z$ can be coupled to $M_z(\mathbf{H} = 0)$ at the microscopic level, where a symmetry-allowed higher-order on-site anisotropy can clearly describe such a property in $Co_{1/3}TaS_2$ (see Supplementary Text and Supplementary Figure 11).

We finally note that a two-fold crystallographic chiral domain related to left or right-handedness, which itself is another interesting topic in $TM_{1/3}MS_2$ (Ref. [35,36]), is deeply linked to the sign relation between $t_z$ and $M_z$; see Supplementary Text. Therefore, the sign relation suggested in Figs. 3 and 4 is limited to one specific chiral domain. However, the presence of the chiral domain is not in conflict with our interpretation of the observed AHE, as it does not affect the measurement of $\sigma_{xy}$ (see Supplementary Text for more explanations).

To summarize, a combination of triangular lattice, chirality, and coplanar 120° spin order gives rise to an interesting feature in $Co_{1/3}TaS_2$: the interplay between the field-tunable toroidal moment and AHE. We have shown that it exhibits large AHE from the vestige of symmetry-protected hourglass Weyl fermions, which has not been reported before. Moreover, $Co_{1/3}TaS_2$ reported in this work is a rare cleavable antiferromagnetic metal: we confirmed that $Co_{1/3}TaS_2$ could be exfoliated down to a few-nm thickness (Supplementary Figure 7). Therefore, it will further offer an exciting chance of studying noncollinear antiferromagnetism and relevant topological properties in a genuine two-dimensional limit, not to mention its rich bulk properties worth further studies.

## Methods

**Sample preparation and characterization.** To obtain $Co_{1/3}TaS_2$ single crystals, we first synthesized polycrystalline $Co_{1/3}TaS_2$ using the solid-state reaction. A mixture of Co (Alfa Aesar, >99.99 %), Ta (Sigma Aldrich, >99.99%), and S (Sigma Aldrich, >99.999%) in a molar ratio of 1.15:3:6 was well ground in an Ar-filled glove box and sintered at 900 ℃ for 10 days in an evacuated quartz ampoule. We measured the powder X-ray diffraction using a commercial machine (Smartlab, Rigaku Japan) to find that it forms in the expected hexagonal structure of the $P6_322$ space group; the corresponding data are shown in Supplementary Figure 1 and Supplementary Table 1. We grew $Co_{1/3}TaS_2$ single-crystal starting with the pre-reacted powder by a chemical vapor transport technique in an evacuated quartz tube with $I_2$ as a transport agent (4.5 mg $I_2$ /$cm^3$). The quartz tube was placed in a two-zone furnace with a temperature gradient from 940 to 860 ℃ for 7~10 days. Finally, we obtained shiny hexagonal crystal pieces (4~13 mm) shown in Fig. 1a.

The obtained single crystals were examined by high-resolution single-crystal x-ray diffraction (XtaLAB PRO, Cu $K_\alpha$, Rigaku Japan) and Raman spectroscopy (XperRam Compact, Nanobase Korea); see Supplementary Figure 2, 3, and Supplementary Table 2. These two measurements confirmed the high quality of our samples (see Supplementary Text for more details). We determined the Co composition ($x$ in $Co_xTaS_2$) using both energy-dispersive X-ray (EDX) spectroscopy and inductively coupled plasma (ICP) spectroscopy (OPTIMA 8300, Perkin-Elmer USA). Both examinations allow us to give the exact value of $x$ within a measurement error close to 1/3 (see Supplementary Table 3). Note that one can slightly tune the resultant stoichiometric ratio by changing excess Co in the starting mixture.

**Bulk property measurements.** Magnetic properties of $Co_{1/3}TaS_2$ were measured by the two commercial magnetometers (MPMS and PPMS-14 with the VSM option, Quantum Design USA). For the temperature-dependent weak ferromagnetic moment (Fig. 2e), we measured the magnetization of the field-cooled (9 T) sample from 2 to 80 K without a magnetic field. Specific heat of $Co_{1/3}TaS_2$ was measured by using the commercial setup (PPMS-14, Quantum design USA). The non-magnetic contribution to the specific heat was estimated by the summation of the Debye model and the electronic contribution linear with the temperature, where we used two Debye temperatures ($\theta_{D1}$= 270 K and $\theta_{D2}$= 611.7 K) for the fitting due to the coexistence of heavy (Ta) and light (S) elements.

Longitudinal and Hall voltage of $Co_{1/3}TaS_2$ were measured by a standard four-probe method using our home-built cryostat, PPMS-9 (Quantum Design, USA), and Teslatron PT (Oxford, UK). All the samples used for the transport measurements were prepared in a straight rectangular shape with a uniform thickness. Hall resistivity data were antisymmetrized to remove any accidental longitudinal signal from the contact misalignment. To observe the temperature dependence of the anomalous Hall effect separated from the ordinary Hall effect (Fig. 2d), we measured the Hall voltage of the field-cooled ($\pm$ 9 T) sample from 2 to 80 K without a magnetic field. We removed the longitudinal component by antisymmetrizing the two datasets measured after the field cool under +9 T and -9 T, respectively.

**Density functional theory (DFT) Calculations.** First-principles calculations were carried out using 'Vienna *ab initio* simulation package (VASP)'[37-39] based on projector augmented wave (PAW) potential[40] and within Perdew-Burke-Ernzerhof (PBE) type of GGA functional[41]. To properly take into account localized Co-3*d* orbitals, we adopted DFT+*U* method[42,43] with the values of $U = 4.1$ eV and $J_{Hund} = 0.8$ eV as obtained by constrained RPA for CoO and Co[44]. We used the Γ-centered $k$-grid of $6 \times 6 \times 4$ for the $\sqrt{3} \times \sqrt{3}$ magnetic supercell. The optimized crystal structure was used with the force criteria of 1 meV/Å. The plane-wave energy cutoff is $500\ eV$. The screw rotation eigenvalues of the Bloch bands were calculated using the 'irvsp'[45]. To investigate the topological properties of electronic states, we obtained Wannier functions using the 'WANNIER90' code[46]. The Berry curvature and anomalous Hall conductivity were calculated using the 'WannierTools' package[47], using the following formula:

$$\Omega_{n,\alpha\beta} = 2\,\mathrm{Im}(\langle \partial_{k_\alpha} u_{nk} | \partial_{k_\beta} u_{nk} \rangle) \tag{2}$$

$$\sigma_{n,\alpha\beta} = \frac{e^2}{(2\pi)^2 h} \sum_n \int d\mathbf{k} f_n(\mathbf{k}) \Omega_{n,\alpha\beta}(\mathbf{k}), \tag{3}$$

where $\Omega_{n,\alpha\beta}$ is Berry curvature, $f_n(k)$ is the Fermi distribution function at 11.6 K, $n$ is the band index, $\alpha$ and $\beta (\neq \alpha)$ denote the Cartesian coordinates ($xyz$), and $u_{nk}$ is the periodic part of the Bloch function. We found it essential to take a dense-enough $k$-grid for accurately estimating the AHC value. After the convergence test, we adopted $301 \times 301 \times 301$.

## Data Availability

The data used in this study are available from the corresponding author upon request.

## Code Availability

Custom codes used in this article are available from the corresponding author upon request.

## Acknowledgments

We acknowledge Dr. S. Lee, and Profs. K-Y Choi, C. Lee, and K. H. Kim for allowing us to use their instruments at the various stages of this work. This work was supported by the Samsung Science & Technology Foundation (Grant No. SSTF-BA2101-05). One of the authors (J.-G.P.) is partly funded by the Leading Researcher Program of the National Research Foundation of Korea (Grant No. 2020R1A3B2079375).

## Author contributions

J.-G.P. initiated and supervised the project. P.P. grew the single crystals and performed all the bulk characterization. P.P. and J.K. measured the transport properties. J.K. performed the mechanical exfoliation and the characterization of the nanoflakes. Y.-G.K. and M.J.H. performed the DFT calculations. P.P., K.H.L., Y.-G.K., M.J.H, H.-J.N., and J.-G.P. contributed to the theoretical analysis and discussion. P.P. and J.-G.P. wrote the manuscript with contributions from all authors.

## Competing Interests

The authors declare no competing financial or non-financial interests.

**Supplementary Information** is available for this paper at [website url].

# Figures

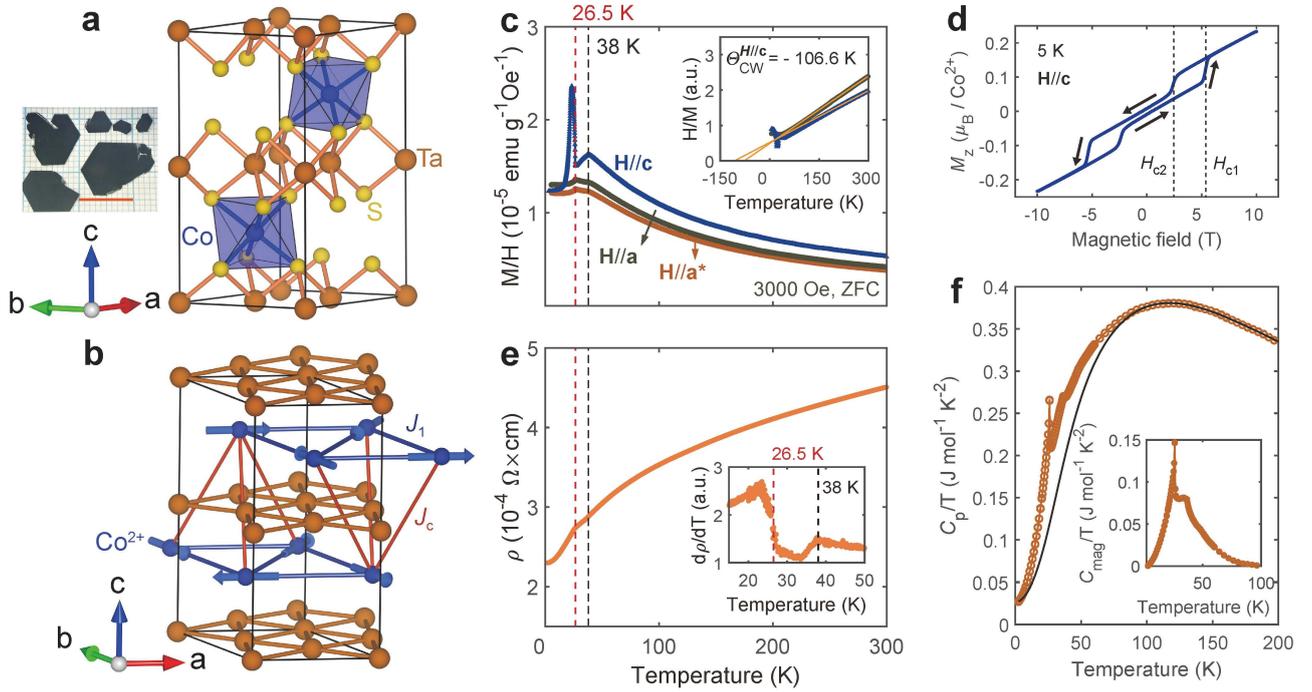

**Fig. 1 | Structure and bulk properties of noncollinear metallic triangular antiferromagnet $Co_{1/3}TaS_2$. a,** A crystallographic unit cell of $Co_{1/3}TaS_2$, where Co atoms are intercalated into octahedral sites. The image shows $Co_{1/3}TaS_2$ single crystals with the solid orange line as long as 10 mm. **b,** Noncollinear magnetic order ($\Gamma_2$) formed by magnetic moments of divalent Co ions. **c,** The magnetization of $Co_{1/3}TaS_2$ as a function of temperature along the three crystallographic axes. The inset shows the Curie-Weiss behavior of $Co_{1/3}TaS_2$. **d,** The field dependence of the magnetization along the c-axis at 5 K. In addition to apparent hysteresis from weak ferromagnetic domains ($\pm H_{c1}$), there is a jump of magnetization at $\pm H_{c2}$. **e,** Temperature-dependent resistivity of $Co_{1/3}TaS_2$ where its first derivative (inset) demonstrates two phase transitions, consistent with the magnetization data. **f,** Specific heat of $Co_{1/3}TaS_2$ as a function of temperature. The inset shows the specific heat after subtracting the non-magnetic contribution (see Methods).

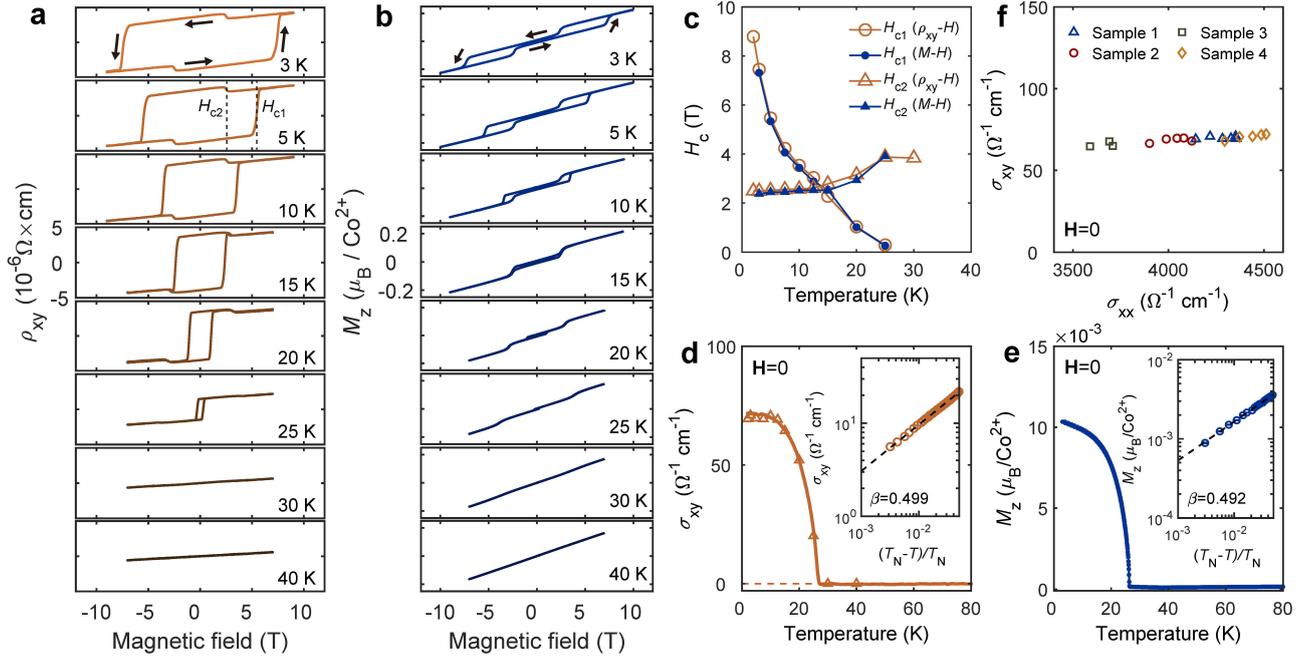

**Fig. 2 | Anomalous Hall effect and its comparison to weak ferromagnetism. a-b,** The field dependence of Hall resistivity $\rho_{xy}$ and magnetization along the c-axis at several temperatures. **c,** Temperature dependence of $H_{c1}$ (circles) and $H_{c2}$ (triangles) determined from the $\rho_{xy}(\mathbf{H})$ (hollow symbols) and $M_z(\mathbf{H})$ (filled symbols) curves. **d-e,** Temperature-dependent zero-field Hall conductivity ($\sigma_{xy}(\mathbf{H}=0)$) and zero-field magnetization along the c-axis ($M_z(H=0)$), measured after a field cooling process (see Methods). Inset shows the critical behavior of $\sigma_{xy}(T,\mathbf{H}=0)$ and $M_z(T,\mathbf{H}=0)$ near 26.5 K and fitted results (black dashed lines) with β=0.499 and 0.492, respectively. **f**, A plot of the longitudinal and zero-field anomalous Hall conductivity for several $Co_xTaS_2$ samples (0.31<x<0.325, see Supplementary Figure 6), which were extracted from the $\rho_{xy}(\mathbf{H})$ curves measured below 15 K.

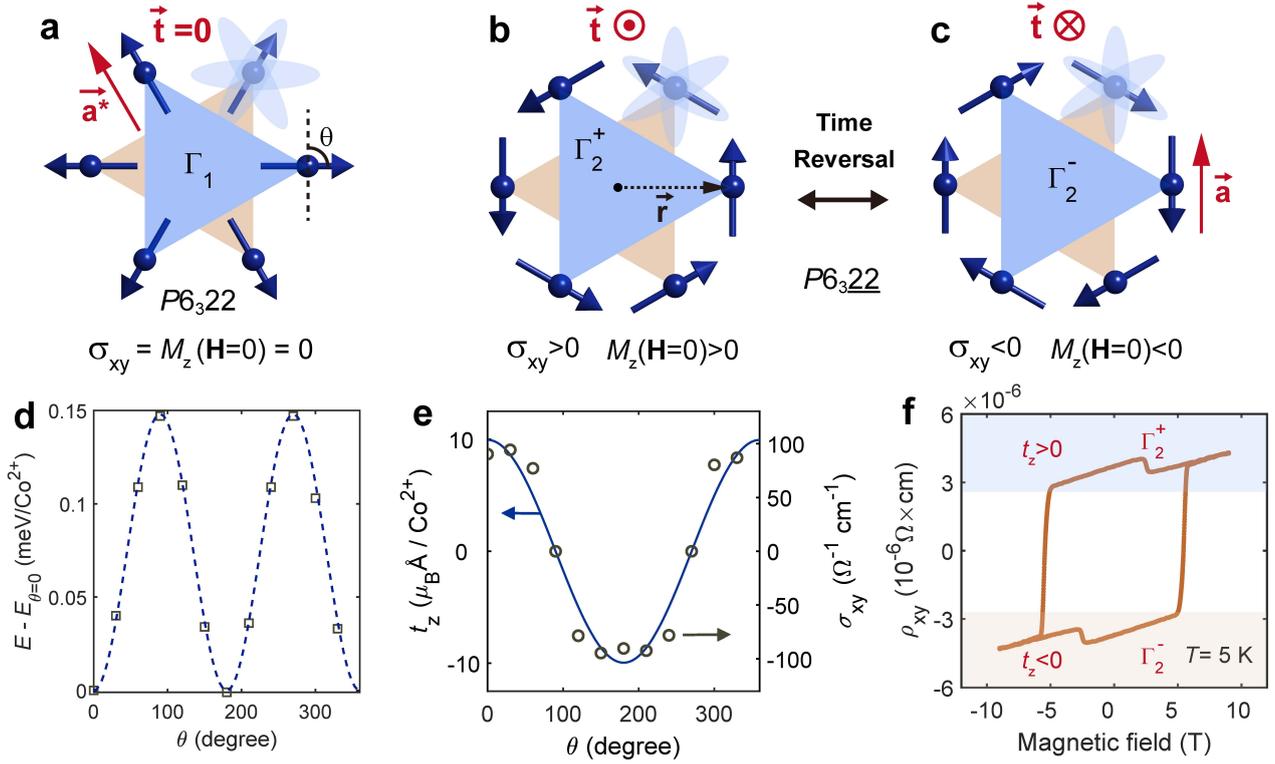

**Fig. 3 | Interplay between a field-tunable toroidal moment and AHE. a,** The $\Gamma_1$ spin configuration having a magnetic space group $P6_322$, in which the symmetry forbids finite $\sigma_{xy}(\mathbf{H}=0)$ and $M_z(\mathbf{H}=0)$. **b-c,** The $\Gamma_2$ spin configuration having a magnetic space group $P6_3\underline{2}\underline{2}$. No symmetry operation forces $\sigma_{xy}$ and $M_z$ to vanish. Three sky-blue ellipses in **a** and **b** denote the three-fold easy-axes to realize the $\Gamma_1$ or $\Gamma_2$ ground state, respectively. **d,** Total energy from the DFT calculation as a function of the uniform spin rotation angle $\theta$ shown in **a**. **e,** Comparison between a toroidal moment along the c-axis (a solid blue line) and calculated $\sigma_{xy}$ for the spin configurations specified by $\theta$ in **a** (black circles). **f,** Interpreting the sign change of AHE as a switch of the toroidal moment. A blue (orange) rectangle denotes a region of the $\Gamma_2^+$ ($\Gamma_2^-$) spin configuration.

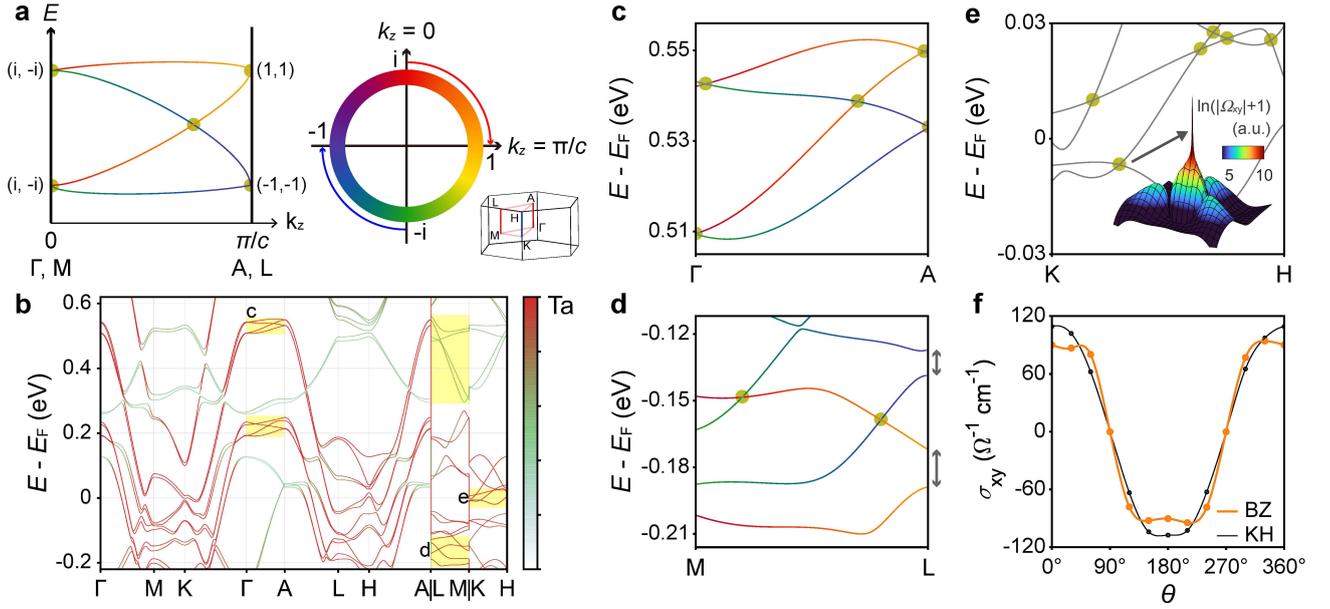

**Fig. 4 | Hourglass Weyl dispersion with noncollinear magnetic order. a,** A schematic band connectivity diagram for a screw-invariant line. The color scale indicates the $2_1$ screw rotation eigenvalues of the Bloch bands. Kramers pairs with eigenvalues $(i, -i)$ are paired at $k_z = 0$ (Γ, M), and those with $(1, 1)$ or $(-1, -1)$ are paired at $k_z = \pi/c$ (A, L). The olive circles represent the Weyl point. The lower right figure shows the Brillouin zone of $Co_{1/3}TaS_2$ with high symmetry points being marked. **b,** The calculated band structure of the $\Gamma_2^+$ spin configuration of $Co_{1/3}TaS_2$. The Ta-$5d$ portions are depicted by color code. Fermi level is set to zero energy. The yellow-colored regions correspond to the hourglass Weyl dispersion, some of which are enlarged in **c**, **d**, and **e**. **c-e,** The calculated band structure along the $\overline{\Gamma A}$, $\overline{ML}$ and $\overline{KH}$ lines, respectively (some of the yellow regions in **b**). Due to the broken TRS, the band degeneracy at the TRIM points is lifted. The color plot in **e** displays the calculated Berry curvature $|\Omega_{xy}|$ near the Weyl point designated by an arrow in a logarithmic scale. **f,** The calculated AHC by integrating over the whole Brillouin zone (BZ) (orange color) and only around the $\overline{KH}$ line (black), respectively. $\theta$ denotes the angle of uniform spin rotation around the c-axis; see Fig. 3a.